\documentclass[twocolumn]{revtex4-1}
\usepackage[utf8]{inputenc}

\date{\today}

\usepackage{graphicx}
\usepackage{systeme,mathtools}
\usepackage{color}

\begin{document}

\title{Improving epidemic testing and containment strategies using machine learning}

\author{Laura Natali\textsuperscript{1}, Saga Helgadottir\textsuperscript{1},  Onofrio M. Maragò\textsuperscript{2}, and Giovanni Volpe\textsuperscript{1} \endgraf
\scriptsize\itshape \textsuperscript{1}Department of Physics,
University of Gothenburg, Sweden \\ \textsuperscript{2}CNR-IPCF, Istituto per i Processi Chimico-Fisici, I-98158 Messina, Italy\\ }

\begin{abstract}
Containment of epidemic outbreaks entails great societal and economic costs.
Cost-effective containment strategies rely on efficiently identifying infected individuals, making the best possible use of the available testing resources.
Therefore, quickly identifying the optimal testing strategy is of critical importance.
Here, we demonstrate that machine learning can be used to identify which individuals are most beneficial to test, automatically and dynamically adapting the testing strategy to the characteristics of the disease outbreak.
Specifically, we simulate an outbreak using the archetypal susceptible-infectious-recovered (SIR) model and we use data about the first confirmed cases to train a neural network that learns to make predictions about the rest of the population.
Using these prediction, we manage to contain the outbreak more effectively and more quickly than with standard approaches.
Furthermore, we demonstrate how this method can be used also when there is a possibility of reinfection (SIRS model) to efficiently eradicate an endemic disease.
\end{abstract}

\maketitle

Compartmental epidemiological models provide a simple and powerful mathematical framework to capture the main features of a disease outbreak in a population \cite{keeling2011modeling, anderson2013population}.
They consider how a disease spreads in a finite population of individuals over a time interval.
The individuals are compartimentalized into categories based on their epidemiological condition.
The first such model, known as the susceptible-infectious-recovered (SIR) model, was proposed in 1927 by Kermack and McKendrick \cite{kermack1927sir} and is still widely employed today \cite{weiss2013sir}.
In the SIR model, there are three categories: susceptible individuals that have never been infected; infectious individuals that are currently infected; and recovered individuals that have previously been infected and are now immunized against the disease.
Initially, all individuals are susceptible except for a limited group of infectious individuals, who seed the disease.

In the event of a disease outbreak, it is often desirable to attempt to contain or eradicate it.
Different factors influence how effective a containment strategy is, including the characteristics of the disease and of the population \cite{Flax2020, maier2020effective}.
However, these characteristics are often difficult to measure or model precisely, especially for novel diseases during their first outbreaks \cite{bi2020epidemiology, carletti2020covid, giordano2020modelling, chinazzi2020effect, perkins2020estimating, bertozzi2020challenges, navascues2020disease, maier2020effective}.
The World Health Organization provides some general guidelines for strategies to prevent disease spread \cite{WHO}, which include travel restrictions, social distancing, and enforced quarantine.
In particular, the isolation of potentially infected individuals is often the most effective measure to limit the spread of the infection.
The safest approach would be to isolate and quarantine all individuals regardless of their epidemiological condition. However, this cannot be implemented and maintained on a large scale for a prolonged period because of its societal and economic deleterious effects \cite{bonaccorsi2020economic}.

In order to implement efficient, cost-effective strategies to contain an outbreak, it is therefore critical to promptly identify infectious individuals.
The most straightforward approach would be to test all the individuals and immediately identify and isolate/treat the infectious ones \cite{lavezzo2020suppression}.
In a real-life large-scale epidemic, however, extensive testing is not usually feasible because of economic and logistic constraints \cite{Aleta2020.05.06.20092841, park2020early, kretzschmar2020impact}.
Therefore, the containment of the disease requires interventions also on individuals who have not been tested yet, which again entails societal and economic costs \cite{ferguson2020report}.

Here, we demonstrate that machine learning can be used to identify an optimized test strategy, i.e., which are the individuals that is most beneficial to test.
Specifically, we introduce a neural-network-powered strategy \cite{goodfellow2016deep, cichos2020machine} for testing and isolating individuals, even though the parameters of the model are not known and infectious individuals can be asymptomatic.
The neural network informs the decision on which individuals should be tested and isolated.
Modelling a disease outbreak using the SIR model \cite{kermack1927sir, weiss2013sir}, we demonstrate that, for an equal number of quarantined individuals, the neural-network-informed strategy manages to contain the disease outbreak more effectively than alternative standard contact-tracing strategies, while autonomously and dynamically adapting to the specifics of the outbreak using only the information about the first confirmed cases.
Furthermore, since for many diseases immunity is not lasting, we also  demonstrate how the neural-network-informed approach can be used to efficiently prevent a new disease from becoming endemic when there is a possibility of reinfection (SIRS model).
We envision that similar methods can be employed in public health to control epidemic outbreaks and to eradicate endemic diseases.

\section*{Results}

\begin{figure*}
\centering
\includegraphics[width=0.9\linewidth]{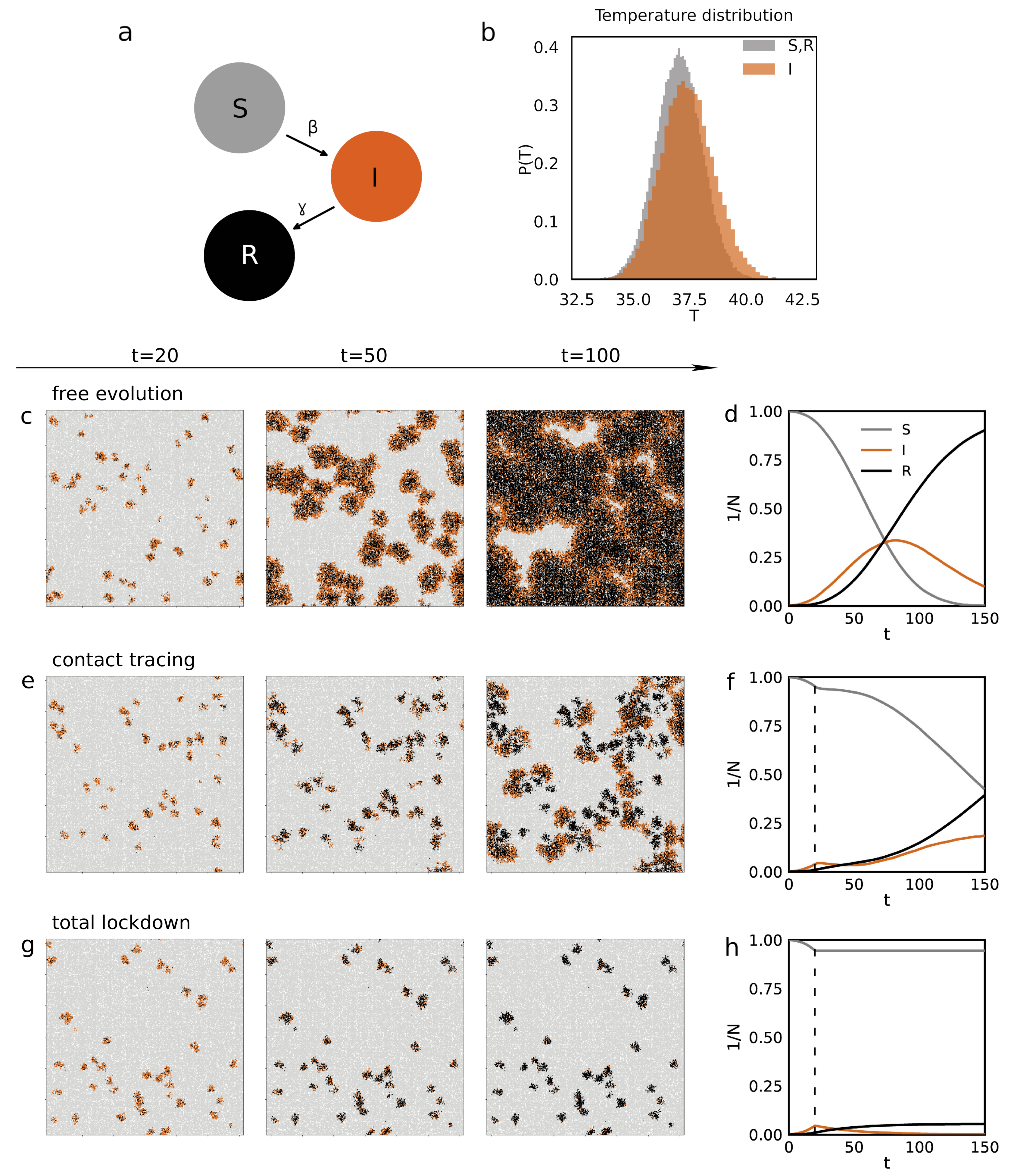}
\caption{{\bf SIR model and containment strategies.}
{\bf a} We consider a population of $10^5$ individuals moving on a square lattice ($320 \times 320$ cells).
Each individual can be either susceptible (S, grey), infectious (I, orange), or recovered (R, black).
At each time step, the susceptible individuals become infectious with probability $\beta$ when they occupy the same cell as an infectious individual, and the infectious individuals recover with probability $\gamma $ becoming immunized against the disease.
{\bf b} Temperature of individuals that are healthy (S and R, $36.8\pm 1.0$) and infectious (I, $37.4 \pm1.2$); note the range of asymptomatic individuals, i.e., infectious individuals with temperature in a healthy range.
{\bf c} Disease spread at times $t = 20, 50, 100$ in the absence of any containment measures and {\bf d} corresponding fraction of the population in each category; unchecked, the disease spreads to almost all the population.
{\bf e} Disease spread using standard contact tracing to isolate potentially infectious individuals starting at $t=20$ (dashed vertical line) and {\bf f} corresponding fraction of the population in each category; the disease spreads more slowly than in {\bf a}, but is not contained.
{\bf g} Disease spread when a total lockdown is implemented at $t=20$ (dashed vertical line) preventing any further spread of the disease and {\bf h} corresponding fraction of the population in each category.
\label{fig:img1}}
\end{figure*}

\subsection*{Epidemic outbreak model and containement strategies}

We model an epidemic outbreak using an agent-based SIR model \cite{keeling2011modeling, black2012stochastic} (see details in Methods, ``SIR model''), where the population consists of $N=10^5$ individuals distributed uniformly on a square lattice with $320 \times 320$ cells, resulting in an average density of $0.98$.
The individuals move as random walkers on the lattice \cite{codling2008random, spitzer2013principles} being each confined to a region with an average radius of $r=10$ cells \cite{ichinose2018reduced}.
All their positions are updated simultaneously at each time step.
Each individual always belongs to one of the SIR categories  (Figure~\ref{fig:img1}a).
At the beginning of the simulation, $50$ individuals ($0.05 \%$ of total population) are randomly selected and made infectious (I).
The rest of the population, instead, is initialized as susceptible (S).
The disease is transmitted with probability $\beta$ when susceptible and infected individuals are occupying the same cell, to mimic the short-range interactions necessary for disease spreading.
An infected individual has a probability $\gamma$ of recovering in each time step, after which it becomes immunized against the disease.
Each individual is also characterized by a ``temperature'', which slightly increases as the disease develops;  the temperature is normally distributed and corresponds to $36.8\pm1.0$ for healthy (i.e., susceptible and recovered) individuals, and to $37.4\pm1.2$ for infectious individuals (Figure~\ref{fig:img1}b), so that there is a significant overlap between the two distributions and, thus, some individuals can be ``asymptomatic''.
We let the model evolve for $150$ time steps, which can be thought of as the days of an epidemic outbreak that lasts approximately six months, but can easily be rescaled to fit another time scale.

Figure~\ref{fig:img1}c provides an example of the \emph{free evolution} of the outbreak in the absence of any containment measures.
By $t=20$, the disease has spread from the initial infectious individuals creating a few hotspots.
These hotspots steadily grow ($t=50$) until most of the population has been infected ($t=100$) and the outbreak starts to subside.
Figure~\ref{fig:img1}d shows how the fraction of individuals in each category varies over time: as the disease spreads, the number of susceptible individuals steadily decreases and the number of recovered ones increases, while the number of infectious individuals initially grows and then slowly decreases until the outbreak ends because essentially the whole population is immunized.

The spread of the disease can be controlled by enacting containment measures.
For example, Figures~\ref{fig:img1}e and \ref{fig:img1}f show the evolution of the outbreak when potentially infectious individuals are isolated based on standard \emph{contact tracing} \cite{ferretti2020quantifying, kretzschmar2020impact, clipman2020rapid, park2020early} (see details in Methods ``Contact tracing'').
At each time step, a fixed number of tests ($N_{\rm test}=100 \ll N$) are performed to assess whether individuals are infectious.
The value of $N_{\rm test}$ is set low enough to simulate a limited access to testing so that only a small portion of the population can be tested (15\% in 150 time steps).
The individuals to be tested are selected randomly from the susceptible individuals with the highest temperature, i.e., those that show more clear symptoms.
Selecting the individuals to be tested in this way presents two advantages compared to a purely random testing strategy: it avoids a slow start (with an initial probability of success around $1/2000$), and it is more representative of reality (where symptomatic cases first indicate an outbreak).
For simplicity, we assume that the test never fails and that there is no delay between performing the test and receiving the result.
However, we remark that the task of identifying the infectious individuals is made harder by the fact that some of their temperatures are in the healthy range (Figure~\ref{fig:img1}b), making them asymptomatic.
The individuals who test positive are quarantined: from that time step on, they neither move nor interact with the rest of the population.
For the tested individuals, the isolation is temporary, so the system knows when they stop being infectious and can safely return to interact with the rest of the population.

Due to the limited number of tests, quarantining only the individuals that test positive is not enough to contain the outbreak.
It is therefore necessary to use contact tracing to isolate also individuals who have not been tested.
(While testing starts from the first time step, contact tracing and isolation of individuals starts only at $t=20$.)
For all detected infectious individuals, we trace back their previous contacts up to $50$ time steps in the past.
Within this group of individuals that interacted with confirmed cases, we test those with the highest temperature.
We rank the other individuals according to their number of contacts with infectious individuals, and, given the same number of contacts, according to their current temperature.
We isolate a number of individuals until reaching a predetermined fraction of the population (here, $25\%$). (See details in Methods ``Contact tracing''.)

It is interesting to compare the free evolution of the outbreak (Figures~\ref{fig:img1}c-d) with the case with isolation based on contact tracing (Figures~\ref{fig:img1}e-f).
While at $t=20$ both outbreaks are similar, the containment measures take hold almost immediately, significantly reducing the size of the outbreaks and the fraction of individuals that are infectious at the same time.
The epidemic outbreak remains confined to a few areas reaching only a part of the population (Figure~\ref{fig:img1}e) and the curve of infected individuals is flatter (Figure~\ref{fig:img1}f).
We remark that, despite its success in slowing down the spread rate of the disease, also the strategy relying on isolation of potentially infectious individuals identified by contact tracing does not lead to a complete suppression of the outbreak, as can be seen from the fact that nearly $20\% $ of the population is infectious still at $t=150$.

Complete eradication of the disease is in principle possible by adopting an unrealistic \emph{total lockdown}, where the whole population is quarantined simultaneously (Figures~\ref{fig:img1}g-h).
From $t=20$, all individuals are isolated so that they cannot move or interact.
Figure \ref{fig:img1}g shows how this leads to an almost immediate containment of the disease hotspots. More interestingly, Figure \ref{fig:img1}h shows how the fraction of infectious individuals quickly drops down and, unlike for the free evolution (Figures~\ref{fig:img1}c-d) and the contact-tracing isolation (Figures~\ref{fig:img1}e-f), reaches zero by $t=120$, so that the disease is extinguished by the end of the simulation.

Different containment strategies lead to different evaluations of the outbreak.
The free evolution (Figure~\ref{fig:img1}c-d) and the total lockdown (Figure~\ref{fig:img1}g-h) approach represent the two limiting policies, leading to the least and the most effective containment.
The contact-tracing isolation (Figure~\ref{fig:img1}e-f) achieves an intermediate level of containment, but does not achieve eradication of the disease, despite isolating up to about $25 \%$ of the population.

\begin{figure*}
\includegraphics[width=\linewidth]{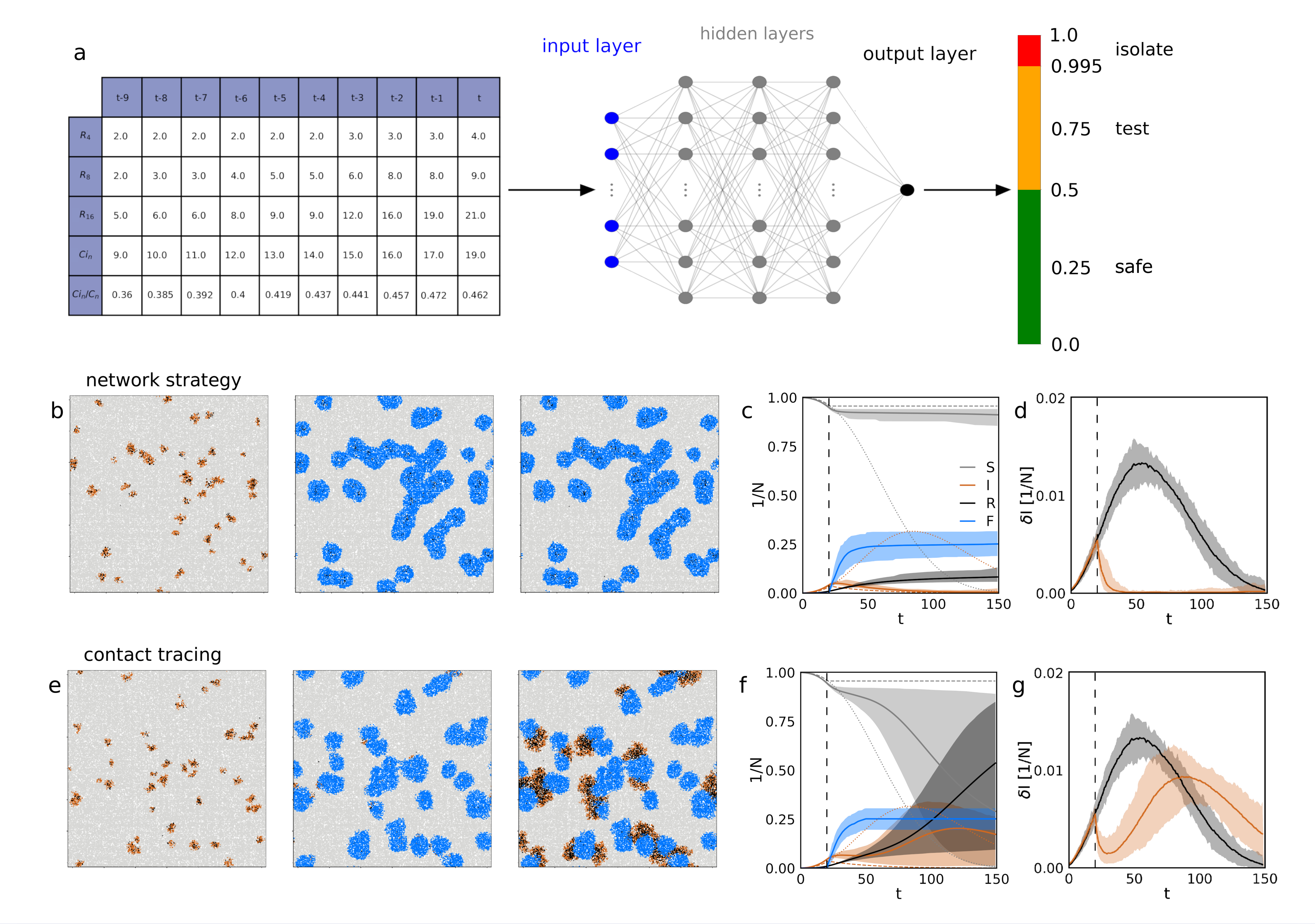}
\caption{{\bf Improved outbreak containment using neural-network-informed testing.}
{\bf a} Structure of the neural network. The inputs (table on the left) are $R_{4,n}(t)$, $R_{8,n}(t)$, $R_{16,n}(t)$, $C^{\rm i}_n(t)$, and $C^{\rm i}_n(t)/C^{\rm tot}_n(t)$, where $R_{r,n}$ is the number of confirmed infectious individuals in a radius $r$ from the individual $n$, $C^{\rm i}_n$ is the number of contacts that individual $n$ has had with confirmed infectious individuals, and $C^{\rm tot}_n$ the total number of contacts for individual $n$; for each parameter, the input includes the history during $10$ time steps ($[t-9, t]$).
The neural network analyzes these inputs through three dense layers and outputs a value $p$ from $0$ (individual predicted to be healthy) to $1$ (individual predicted to be infectious): individuals with $p>0.995$ are directly quarantined, and individuals with $p\in[0.5, 0.995]$ are tested starting from individuals with the highest temperatures until the depletion of the available tests.
{\bf b}-{\bf d} Disease evolution when the testing and isolation strategy is determined based on the output from a neural network:
{\bf b} Snapshots of susceptibles (S, grey), infectious (I, orange), recovered (R, black), and frozen (F, blue) individuals at time steps $ t = 20, 50, 100$.
{\bf c} Corresponding fraction of the population in each category compared with the two limiting cases of free evolution (dotted orange line, see also Figures~\ref{fig:img1}c-d) and full lockdown (dashed orange line, see also Figures~\ref{fig:img1}g-h). The isolation of individuals starts at $t=20$ (dashed black line).
The solid lines indicate the average over $N_{\rm runs}=100$, while the shaded areas correspond to $90\%$ confidence interval.
{\bf d} The number of new cases $\delta I$ for the neural-network-informed testing (orange line) compared to the free evolution (black line).
{\bf e}-{\bf g} Comparison to a standard contact-tracing strategy (see Figures~\ref{fig:img2}e-f) where the same number of individuals are quarantined as in {\bf b}-{\bf d} where we employ the neural-network-informed strategy: the number of infectious individuals and the spread of the disease are greatly reduced when employing the  neural-network-informed strategy.
\label{fig:img2}}
\end{figure*}

\subsection*{Neural-network-informed testing}

It would be desirable to achieve disease eradication as in the total-lockdown strategy (see Figures~\ref{fig:img1}g-h), but isolating only part of the population as in the contact-tracing strategy (see Figures~\ref{fig:img1}e-f).
To achieve this, we propose a strategy that employs a neural network to inform which individuals to test and isolate.

The schematic of the neural network we employ is shown in Figure~\ref{fig:img2}a (see details in Methods ``Neural network'').
In general, a neural network receives some inputs, elaborates them through of a series of hidden layers of artificial neurons, and returns an output \cite{mehlig2019artificial}.
In our case, the input consists of contact-tracing information for a given individual $n$ for the last $10$ time steps.
Specifically, we provide the neural network with five time series: $R_{4,n}(t)$, $R_{8,n}(t)$, $R_{16,n}(t)$, $C^{\rm i}_n(t)$, and $C^{\rm i}_n/C^{\rm tot}_n(t)$. The first three indicate the number of tested infectious individuals within a distance $r=4$, $8$, and $16$ cells from the considered one. $C^{\rm tot}_n(t)$ is the total number of contacts (i.e., defined as individuals occupying the same cell at the same time) and $C^{\rm i}_n(t)$ is the number of contacts with confirmed infectious individuals.
Then, the neural network elaborates this information through three dense layers of artificial neurons.
Finally, the neural network outputs a value $p$, representing the risk of being infectious at the current time step, between $0$ for a putatively healthy individual and $1$ for a putatively infectious individual.
Individuals with $p>0.995$ are immediately isolated, while individuals with $p\in[0.5, 0.995]$ are slated to be tested, starting from the individuals with the highest temperatures until the depletion of all available tests.
In this way, we manage to freeze the infectious individuals that are easy to identify, while optimizing the deployment of the available tests: we use the tests principally to achieve a better understanding of the extent and distribution of the disease.

Neural networks are supervised machine learning methods and, therefore, require training \cite{mehlig2019artificial}.
In general, the training of a neural network is performed by providing the neural networks with a series of inputs and corresponding known outputs \cite{mehlig2019artificial}.
In our case, we can only use for training individuals that have already been tested within each run of the simulation (see details in Methods ``Neural-network training''). Therefore, we start training at $t=20$, when we have tested $2000$ individuals.
In subsequent time steps, the size and accuracy of the training data set increases with the number of performed tests, so we repeatedly retrain the neural network to improve its performance.
This leads to a positive feedback loop, where a better-trained neural network selects more efficiently individuals for testing, which in turn provides better insights into the disease distribution, which finally improves the training data set available to further improve the performance of the neural network.

Figure \ref{fig:img2}b depicts the snapshots of the system at $t=20,50,100$. The color code is the same as that used in  Figure~\ref{fig:img1}, with the addition of frozen individuals (F) indicated in light blue.
Until $t=20$, the outbreak evolves freely, analogously to Figure~\ref{fig:img1}c, while enough data are accumulated to train the neural network.
From $t=20$ and onward, the neural-network predictions are used to inform which individuals to isolate and test.
By $t=50$, all outbreaks have been identified and surrounded by frozen individuals.
Subsequently ($t=100$), the outbreaks remain under control and are prevented from spreading, in stark contrast with the wide spread of the disease in free evolution ($t=100$ in Figure~\ref{fig:img1}c).

The orange solid line in Figure~\ref{fig:img2}c shows the fraction of the population that is infectious as a function of time.
Shortly after we switch on the neural network ($t=20$), the infectious fraction reaches its maximum ($5.1 \%$ at $t=26$) and subsequently rapidly decreases to zero.
Correspondingly, the number of recovered (black solid line) and susceptible (gray solid line) individuals reach a plateau.
In particular, the fraction of individuals that are infected and eventually recover is $8\pm4\%$.

The number of frozen individuals is initially zero and quickly increases in the first stages of neural-network-informed testing, eventually reaching the set value of $25\%$ of the total population.
We can compare the curve of the infectious individuals using the neural-network-informed testing and isolation (orange solid line) with the limiting cases of free evolution (orange dotted line, cfr. Figure~\ref{fig:img1}c) and of total lockdown (orange dashed line, cfr. Figure~\ref{fig:img1}g).
By isolating only $25\%$ of the population, the neural-network-informed strategy achieves a containment of the epidemic similar to that achieved by the full lockdown.

Figure~\ref{fig:img2}d represents the fraction of new infectious individuals per time step for the neural-network-informed strategy (orange line) and for the free evolution of the epidemics (black line).
The free-evolution curve reaches a maximum at $t=59$ corresponding to $\delta I(59)=1.4 \pm 0.2\%$.
The curve for the neural-network-informed strategy starts decreasing immediately after isolation starts at $t=20$, corresponding to a peak value $\delta I(20) = 0.55 \pm 0.08 \%$, and stably reaches zero around $t= 50$.

Figures~\ref{fig:img2}e-g provide comparisons with a standard contact-tracing strategy, where the same number of individuals are tested and isolated as described in detail in the previous section.
Figure~\ref{fig:img2}e shows snapshots of the system at $t=20,50,100$:
starting from the same number of hotspots ($t=20$), contact tracing manages to identify all regions reached by the disease ($t=50$), but the disease can still spread due to the limited number of individuals that can be isolated ($t=100$).
Figure~\ref{fig:img2}f shows that, differently from the case of the neural-network-informed strategy (Figure~\ref{fig:img2}c), the increase of the fraction of infected individuals slows down for some time steps, but then starts again to grow reaching a peak at $t=120$ corresponding to about $20\%$ of the total population.
The total number that have been infected at the end of the simulation (i.e., all infectious and recovered individuals at $t=150$) is strikingly lower for the neural-network-informed strategy ($6\%$ to $14\%$) than for the contact-tracing-based strategy ($30\%$ to $89\%$).
The wide shaded area in Figure~\ref{fig:img2}f is nearly $7$ times larger than in Figure~\ref{fig:img2}c, showing that the contact tracing is less stable against different evolution patterns of an epidemic with same underlying SIR parameters.
The orange line in Figure~\ref{fig:img2}g shows the fraction of new infectious individuals $\delta I$ as a function of time, which is non-zero at the end of the simulation, unlike for the neural-network-informed strategy (orange line in Figure \ref{fig:img2}d). We can therefore conclude that contact tracing is less effective than the neural network for the same number of frozen individuals.

\begin{figure*}
\centering
\includegraphics[width=\linewidth]{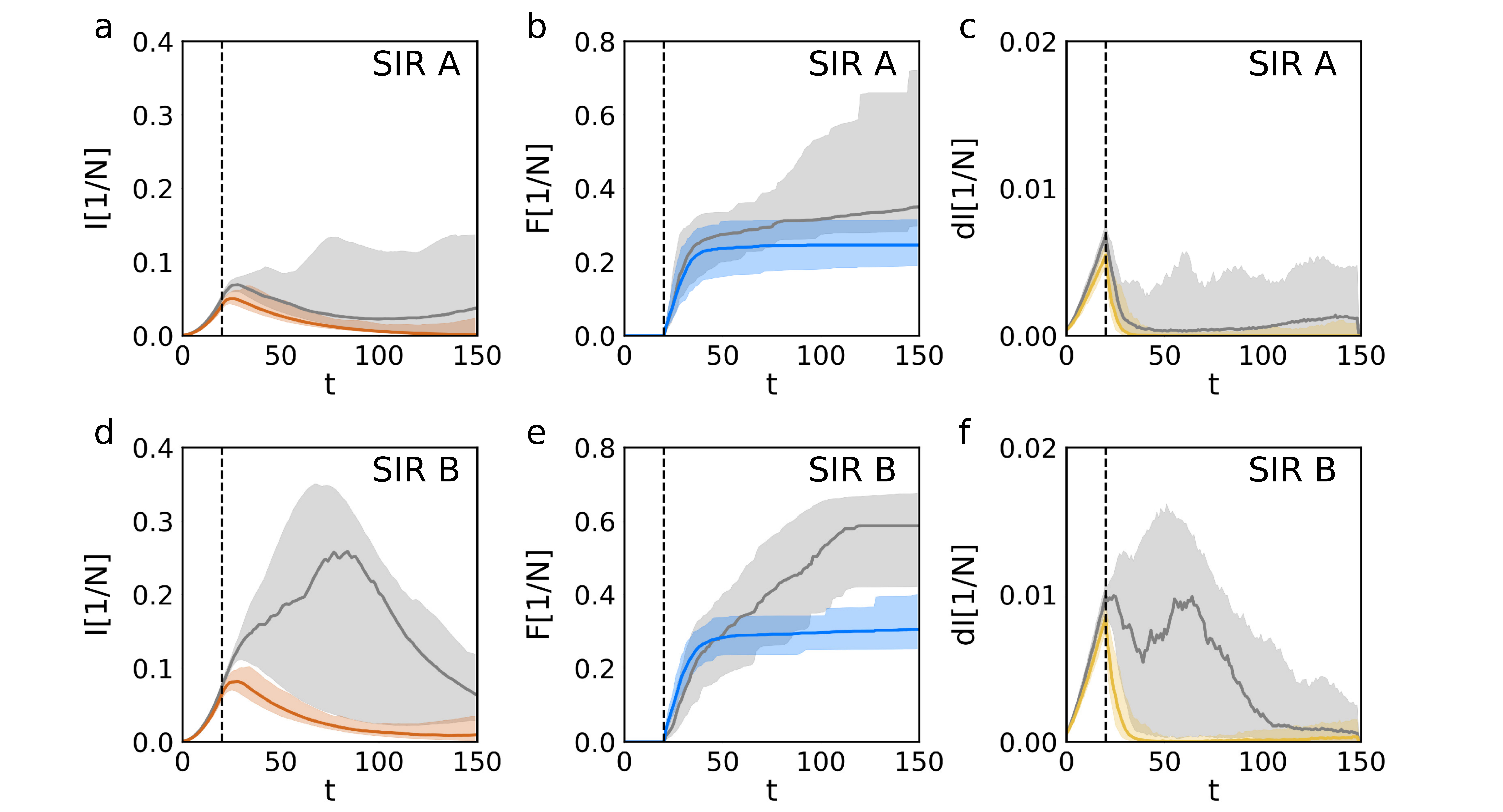}
\caption{{\bf Automatic and dynamic adaptation of the neural network to the underlying SIR parameters.}
Two independent neural networks (NN$_{\rm A}$ and NN$_{\rm B}$) are trained for two SIR models (SIR$_{\rm A}$ and SIR$_{\rm B}$, respectively) with different parameters ($\beta_{\rm A}=0.6$ and $\gamma_{\rm A}=0.03$ for SIR$_{\rm A}$, and $\beta_{\rm B}=0.8$ and $\gamma_{\rm B}=0.03$ for SIR$_{\rm B}$).
SIR$_{\rm A}$ and NN$_{\rm A}$ are the same as those employed in Figures~\ref{fig:img2}b-d.
{\bf a} Infectious individuals (orange line), {\bf b} frozen individuals (blue line), and {\bf c} new infections (yellow line) when NN$_{\rm A}$ is used on SIR$_{\rm A}$.
The gray lines are the corresponding curves when NN$_{\rm B}$ is used instead, showing a clear decrease in  performance.
{\bf d}-{\bf f} Corresponding plots where NN$_{\rm B}$ (colored lines) and NN$_{\rm A}$ (gray lines) are used on SIR$_{\rm B}$.
Overall, these results show that the neural network gets automatically optimized for the parameters of the underlying outbreak.
In all cases, the shaded areas represent the $90\%$ confidence intervals obtained from 100 simulations.
\label{fig:img3}}
\end{figure*}

\subsection*{Automatic and dynamic adaptation to the outbreak characteristics}

An important characteristic of the neural-network-informed strategy is that it can automatically and dynamically adapt itself to the underlying characteristics of the outbreak.
In our model, this means that the neural network does not need to have explicit knowledge of the underlying SIR model.
More generally, the neural network can adapt to other kinds of outbreaks and also take into account the effects of the containment measure put in place.

Figure~\ref{fig:img3} demonstrates the ability of the neural-network-informed strategy to automatically and dynamically adapt itself to the underlying characteristics of the outbreak.
The colored solid lines in Figures~\ref{fig:img3}a-c reproduce the performance of the strategy presented in Figures~\ref{fig:img2}b-d, which is informed by a neural network (NN$_{\rm A}$) trained on the data obtained from an outbreak (SIR$_{\rm A}$, $\beta_{\rm A}=0.6$ and $\gamma_{\rm A}=0.03$), in terms of the evolution of infectious individuals (orange line, Figure~\ref{fig:img3}a), frozen individuals (blue line, Figure~\ref{fig:img3}b), and new infections in each timestep (yellow line, Figure~\ref{fig:img3}c).
We then apply NN$_{\rm B}$, i.e., another neural network trained on a different outbreak whose underlying SIR model has a slighlty different transmission rate (SIR$_{\rm B}$, $\beta_{\rm B}=0.8$ and $\gamma_{\rm B}=0.03$).
The resulting performance can be seen in the gray lines in Figures~\ref{fig:img3}a-c.
While overall NN$_{\rm B}$ manages to improve the outbreak with underlying SIR$_A$ model compared to its free evolution, it performs much worse that NN$_{\rm A}$. At the end of the simulation in Figures~\ref{fig:img3}a, the fraction of infectious individuals is still in the range ($0.12 \% $ to $ 13.7\%$) of the population for the gray confidence bands, while the overall fraction of individuals in isolation is in the range ($30\%$ to $72\%$), as shown in Figures~\ref{fig:img3}b.
This suggests that, thanks to its training using the information acquired by the testing during the first 20 time steps, the neural-network-informed strategy gets fine-tuned to the specific characteristics of the underlying outbreak.

We further validate the fine-tuning of the neural network by training NN$_{\rm B}$ on the testing data obtained from the outbreak with underlying model SIR$_{\rm B}$.
The colored lines in Figures~\ref{fig:img3}d-f show the results of applying NN$_{\rm B}$ on the SIR$_{\rm B}$ outbreak, which demonstrate a good containment of the outbreak.
Instead, the gray lines show what happens when using NN$_{\rm A}$, which leads to a much worse outcome. In this scenario, the peak for the curve of infected is around $t=84$ and $25.7\%$ against $8.1 \%$ of the population for the training performed on SIR$_{\rm B}$. Figures~\ref{fig:img3}f shows that $\delta$I oscillates between $540$ and $995$ new cases per time step in the interval $t \in [20,73]$ before decreasing.

\begin{figure*}
\centering
\includegraphics[width=\linewidth]{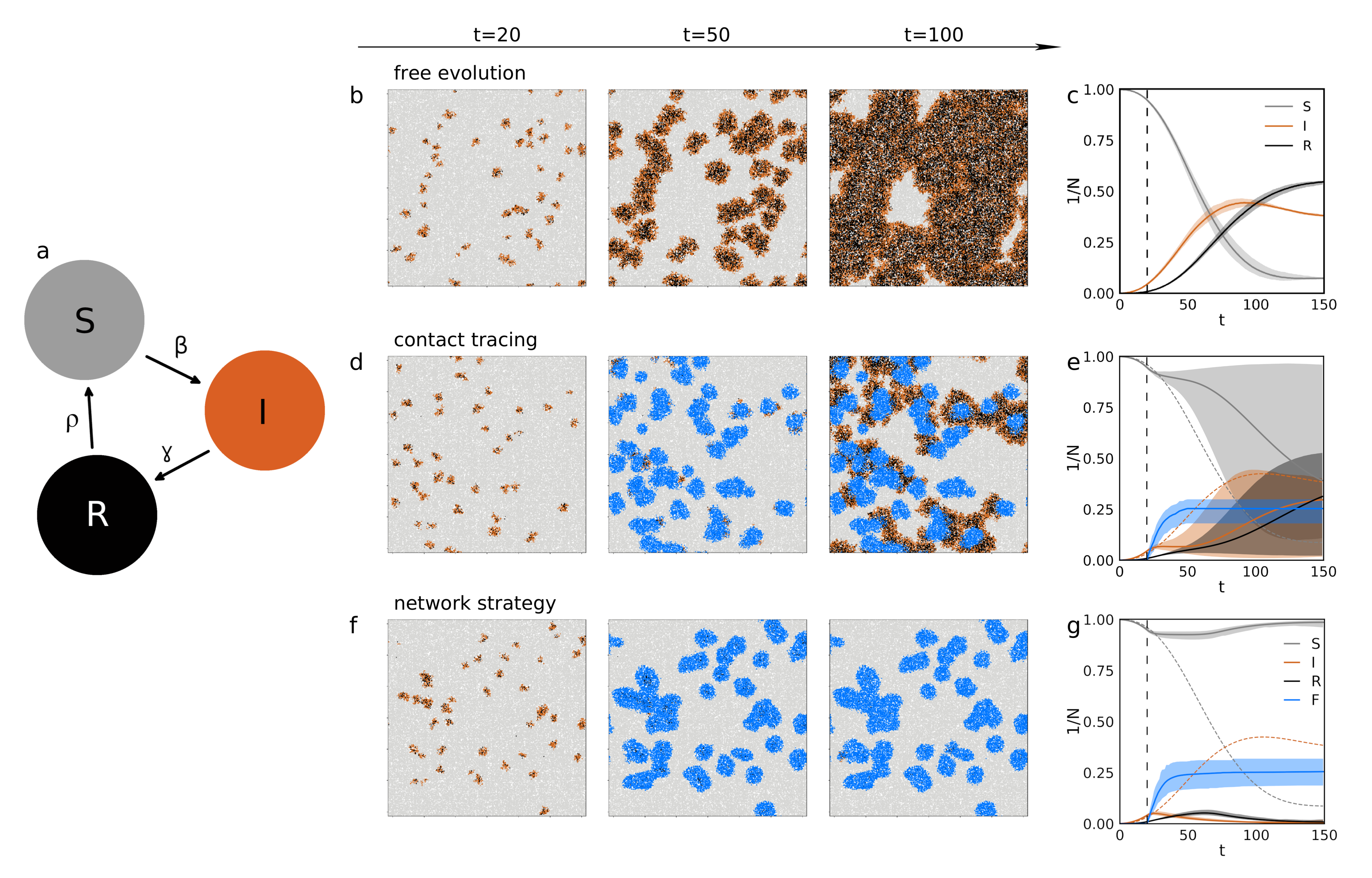}
\caption{
{\bf SIR model with temporary immunity (SIRS).}
{\bf a} We consider a model with possibility of reinfection (SIRS model), where at each time step, recovered individuals have a probability $\rho$ of becoming susceptible again (specifically, we use $\beta = 0.06$, $\gamma = 0.03$, and $\rho = 0.02$).
Without countermeasures, this leads to an endemization of the disease.
{\bf b} Disease spread at $t = 20, 50, 100$ in the absence of any containment measures and
{\bf c} corresponding fraction of the population in the susceptible (S, gray line), infectious (I, orange line) and recovered (R, black line) categories;
the endemization can be inferred from the stabilization of the fraction of infectious individuals towards the end of the simulation.
{\bf d} Disease spread using standard contact tracing to isolate potentially infectious individuals starting at $t = 20$ (dashed vertical line in {\bf f}) and
{\bf e} corresponding fraction of the population in each category, including frozen individuals (F, blue line); while the disease spreads less, it still becomes endemic.
{\bf f} Disease spread when employing a neural network to inform testing starting at $t = 20$ (dashed vertical line in {\bf g}) and
{\bf g} corresponding fraction of the population in each category; in this case, the disease is completely eradicated.
The dashed lines in {\bf e} and {\bf g} correspond to the free evolution of the disease and are reported from {\bf c} for comparison.
In all cases, the solid lines indicate the average over multiple runs ($N_{\rm runs}=100$), while the shaded areas correspond to $90\%$ confidence intervals.
\label{fig:img4}}
\end{figure*}

\subsection*{Disease eradication with possibility of reinfection}

We now consider the case when the immunity against the disease is not permanent \cite{long2020clinical, seow2020longitudinal,shaman2020will}.
Thus, we consider a SIRS model (Figure~\ref{fig:img4}a), which is an extension of the SIR model where recovered individuals have a probability $\rho$ at each time step to become again susceptible \cite{long2020clinical,seow2020longitudinal} (see details in Methods ``SIRS model'').
In the absence of any containment measures, the possibility of reinfection leads to an endemization of the disease.
Figure~\ref{fig:img4}b shows such free evolution of the disease:
from the initial hotspots ($t = 20$), the disease spreads quickly to a large portion of the population ($t = 50$) until reaching a steady state.
Figure~\ref{fig:img4}c shows how the fraction of individuals in each category varies over time: during the initial spread of the disease, the number of susceptible individuals steadily decreases and the number of infectious ones increases;
once the disease reaches its steady state, the fraction of infectious individuals stabilizes to a value that depends on the characteristics of the SIRS model, i.e., on the value of its parameters $\beta, \gamma$ and $\rho$.
Therefore, the disease becomes endemic \cite{keeling2011modeling}.

Figure~\ref{fig:img4}d-e show the development of the disease when a standard contact-tracing-based containment strategy is implemented, like that employed in Figures~\ref{fig:img1}e-f.
The solid lines represents the averages for susceptibles (S, gray), infectious (I, orange), recovered (R, black) and frozen (F, blue) individuals throughout the simulation.
The color bands, which denote the $90\%$ confidence interval, is  larger than those in Figure~\ref{fig:img2}f;
this implies that the performance of the contact-tracing strategy can vary significantly depending on the specific outbreak.
It can be seen that this containment approach manages to reduce the number of infectious individuals in the steady state of the disease, but not to eradicate the disease itself.

Finally, Figure~\ref{fig:img4}f-g show the performance of the neural-network-informed strategy. We employ the same approach and neural network architecture shown in Figure~\ref{fig:img2}a and the same strategy that we employed to contain the outbreaks in the SIR model shown in Figures~\ref{fig:img2}b-d.
Briefly, we start testing individuals from the beginning of the simulation accumulating data to train the neural network.
From $t = 20$, we start training the neural network to predict infectious individuals and use this information to decide which individuals to isolate and test.
The neural-network-informed strategy manages to eradicate the disease, as can be seen from the fact that the fraction of infectious individuals approaches zero by the end of the simulation (orange solid line in Figure~\ref{fig:img4}g), while the number of susceptible individuals increases as recovered individuals gradually lose their immunity.
Therefore, by employing the neural-network-informed strategy, it is possible to prevent the initial outbreak from leading to the endemization of the disease.

\section*{Discussion}

The current outbreak of the novel coronavirus disease (COVID-19) \cite{zhu2020novel, wu2020new, world2020coronavirus, li2020emergence, bi2020epidemiology} has dramatically brought to worldwide attention the crucial importance of epidemiological models for choosing the best strategies and policies to contain disease outbreaks \cite{ferguson2020report, bi2020epidemiology, maier2020effective, navascues2020disease, giordano2020modelling}.
Machine-learning approaches have been already proposed to help disease diagnosis  \cite{pina2020virtual} and  epidemics handling \cite{navascues2020disease}.
In fact, in the last few years, various neural-network architectures have been employed to manage human diseases  \cite{wang2020covidnet,melin2020multiple, lalmuanawma2020applications,dandekar2020neural}, such as malaria \cite{kiang2006meteorological}, and animal diseases, such as in swine flu  \cite{AUGUSTA2019187}.
In this work, we have now shown how a neural-network-informed strategy can improve the containment of an epidemic, even when only a small number of specific tests is available and some of the individuals are asymptomatic.
This improvement can be seen in three key aspects.
First, integrating the neural network into the outbreak handling improves the performance of contact tracing, while performing the same number of tests and isolating the same fraction of individuals.
Second, the neural network autonomously tunes its weights to the ongoing outbreak, without needing to explicitly know its underlying model or its parameters, and therefore does not require a priori knowledge of the disease outbreak characteristics.
Third, since the neural network is regularly retrained as new data become available, it can automatically and dynamically adapt itself to the evolution of the outbreak as well as to the changes in the behavior of the population, e.g., due to containment measures or different social habits.
As a striking example, we have shown that, in the case of temporary immunization, the neural-network-informed strategy can prevent a disease outbreak from becoming endemic.

Even though we used a SIR model to describe the dynamics underlying the disease, the neural network will automatically adapt itself to different underlying dynamics described by more complex epidemiological models, which might include, e.g., the disease incubation time \cite{giordano2020modelling}, delays in the testing process \cite{kretzschmar2020impact}, or even different patterns of movement of the individuals (e.g., periodic motion, and long-range travel) \cite{chinazzi2020effect}.
It is also possible to provide the neural network demographic information (e.g., individual risk factors, such as age, employment, and preexisting conditions) as well as spatial information (e.g., the location of the individuals, differentiating various places of aggregation, such us hospitals, markets, and schools), or even simple access medical tests (e.g. cough recordings \cite{laguarta2020covid}).
Furthermore, the neural-network-informed approach presented in this work can be generalized to other situations, such as fire prevention \cite{tonini2020machine} or econometrics \cite{athey2019machine}.

\section*{Methods}

\subsection*{SIR model}

We divide the population of $N=10^5$ identical individuals into three epidemiological categories: susceptible individuals $S$, infectious individuals $I$, and recovered individuals $R$, as in the original SIR model \cite{kermack1927sir}.
The individuals move on a square lattice with side $l=320$ according to a stochastic model \cite{berg1993random, codling2008random}.
The position of each individual $n\in[1,N]$ at each time step $t\in[0,150]$ is given by its coordinates $\textbf{x}_n(t)=[x_n(t),y_n(t)]$.
Each individual is an independent random walker confined to move within a small area of the lattice centered around its initial random position $\textbf{x}_n(0)=[x_n(0),y_n(0)]$. At each time step, it can move in its Moore neighborhood \cite{seitz2012natural} according to the following displacements:
\begin{equation}
    \Delta x_n
    =
    \begin{cases}
        -1 \quad & \mbox{with probability } {1\over3} - k [x_n(t) - x_n(0)] \\
        0 \quad & \mbox{with probability } {1\over3} \\
        +1 \quad & \mbox{with probability } {1\over3} + k [x_n(t) - x_n(0)]
    \end{cases}
\end{equation}
\begin{equation}
    \Delta y_n
    =
    \begin{cases}
        -1 \quad & \mbox{with probability } {1\over3} - k [y_n(t) - y_n(0)] \\
        0 \quad & \mbox{with probability } {1\over3} \\
        +1 \quad & \mbox{with probability } {1\over3} + k [y_n(t) - y_n(0)]
    \end{cases}
\end{equation}
where $k=0.04$ determines the radius $r_k\approx10$ cells within which each individual moves.
The positions of all individuals are updated synchronously and independently from each other.

The spread of the infection occurs because when a susceptible individual occupies the same cell as an infectious individual, it becomes infectious with probability $\beta$ in each time step.
The transmission applies only for the infectious individuals that are not frozen.
Each infectious individual becomes recovered with probability $\gamma$ at each time step.
The parameters used are $\beta=0.6$ and $\gamma =0.03$, except for Figure \ref{fig:img3}, where we also employ $\beta=0.8$.

Each individual is also characterized by a ``temperature'', which is normally distributed and corresponds to $36.8\pm1.0$ for healthy (i.e., susceptible and recovered) individuals, and to $37.4\pm1.2$ for infectious individuals, with a great overlap between the two distributions (Figure~\ref{fig:img1}b).

\subsection*{SIRS model}

The SIRS is an alternative to the SIR model that assumes the immunization to the disease is temporary.
Therefore, recovered individuals lose immunization and return susceptible with probability $\rho$ in each time step.
We employ $\rho = 0.02$.

\subsection*{Contact tracing}

We keep track of individuals that occupy the same cell at a certain time step by introducing the contact matrix:
\begin{equation}
    c_{nm}(t)
    = \delta({\bf x}_n(t)-{\bf x}_m(t)),
\end{equation}
where $\delta$ is the Kronecker delta, which has value 1 if the pair of individuals $n$ and $m$ occupy the same cell at time $t$, and 0 otherwise.
Thus, the total number of contacts for individual $n$ for the $50$ time steps before time $t$ is
\begin{equation}
    C^{\rm tot}_n(t)
    =
    \sum_{\tau=t-50}^{t}
    \sum_{m \neq n} c_{nm}(\tau).
    \label{eq:Ctot}
\end{equation}
The number of contacts with confirmed infectious individuals is
\begin{equation}
    C^{\rm i}_n(t) = \sum_{\tau=t-50}^{t}
    \sum_{m \neq n}
    c_{nm}(\tau)
    \delta^{\rm i}_m(t),
    \label{eq:ci}
\end{equation}
where $\delta^{\rm i}_m(t)$ is 1 if individual $m$ has already been tested and found positive at time $t$, and 0 otherwise.
When implementing the lockdown strategy based on contact tracing, we list the agents in descending order as a function of $C^{\rm i}_n(t)$, and we sort those with equal value based on their temperature.
At each time step, we select for testing the first $N_{\rm test}=100$ individuals in this list.
We use the rest of such list for selecting individuals to freeze, whose number is set to match that of the neural-network-informed strategy.
In this way, we can compare the two approaches using the same number of tests and the same number of frozen individuals.
When the target number of individuals to isolate is larger than the individuals in the contact list (e.g., at the beginning of the simulation when the number of confirmed cases is small), we build an additional list from where to select the remaining individuals, which includes individuals that never had direct interactions with confirmed cases, but have been within a radius of 8 cells in the last 50 time steps; we sort also this additional list based on the temperature of the individuals.

\subsection*{Neural network}

We employ a dense neural network with three hidden layers with $16$ neurons each and ReLU activation function \cite{agostinelli2014learning, book_NN}.
The output layer has one single neuron with a softmax activation function returning a value $p \in [0,1]$.
Additionally, we use dropouts for the hidden layers as a way to avoid overfitting \cite{srivastava2014dropout} (dropout rate $0.2$, so that in each training epoch only $80\% $ of the neurons is activated).

The input to the neural network at time $t$ includes $R_{4,n}(t)$, $R_{8,n}(t)$, $R_{16,n}(t)$, $C^{\rm i}_{n}(t)$, and $C^{\rm i}_{n}/C^{\rm tot}_{n}(t)$ for time steps $[t-9, t]$, where $C^{\rm i}_{n}(t)$ and $C^{\rm tot}_{n}(t)$ are the number of infectious and total contacts (Eqs.~\ref{eq:Ctot} and \ref{eq:ci}), and $R_{r,n}(t)$ is the number of individuals that have tested positive within a radius $r$:
\begin{equation}
    R_{r,n}(t)
    =
    \sum_i \delta \left(
        r
        -
        \parallel \mathbf{x}_n(t) - \mathbf{x}_i(t) \parallel
    \right),
\end{equation}
where the summation is over all infected individuals.

The training of the neural network is performed using information relative to the individuals that have already been tested (which is split between a training set and a validation set \cite{krogh1995neural}).
The loss function is the mean square error, we use the stochastic gradient descent method implemented in the Adam optimizer \cite{kingma2014adam, ruder2016overview}, and the number of training epochs is fixed to 100 (see Supplementary Figure S1).
While we use only two labels for the training ($0$ for susceptible individuals and $1$ for infectious individuals), the trained network returns a prediction that is a continuous value $p \in [0, 1]$.

Using the prediction of the network, we split the individuals that have not been tested yet into three groups:
(1) $p>0.995$: individuals with a high chance of being infectious, who are frozen without testing.
(2) $0.5<p<0.995$: individuals with a medium chance of being infectious, amongst which the $N_{\rm test} = 100$ individuals with the highest temperature are tested.
(3) $p<0.5$: individuals with a low chance of infection.

We implement the neural network using the Python libraries Tensorflow and Keras \cite{chollet2015keras}.

\section*{Acknowledgement}
We acknowledge support from the MSCA-ITN-ETN project \emph{ActiveMatter} sponsored by the European Commission (Horizon 2020, Project Number 812780).


\end{document}